\documentclass[sigplan,9pt,screen,natbib,nonacm]{acmart}
\pdfoutput=1
\setcopyright{none}
\usepackage{mdwtab}
\usepackage{mdwlist}
\usepackage{amsmath}
\usepackage{comment}
\usepackage{url}
\usepackage{hyphenat}

\usepackage{listings}

\lstdefinestyle{sOcaml}{language=[Objective]Caml,
  morekeywords={effect,perform,locus},
  literate={+}{{$+\:$}}1 {/}{{$/$}}1 
           {=}{{$=$}}1
           {>}{{$>$}}1 {<}{{$<$}}1
           {<>}{$\not=$}1
           {->}{{$\rightarrow$}}2 {>=}{{$\geq$}}2 {<-}{{$\leftarrow$}}2
           {<=}{{$\leq$}}2
           {==>}{{$\mapsto$}}2
           {|}{{$\mid$}}1
           {|>}{{$\triangleright$}}1
           {>>}{{$\rhd$}}1
           {'a}{$\alpha$}1
           {'b}{$\beta$}1
           {'c}{$\gamma$}1
           {'e}{$\epsilon$}1
           {'n}{$\nu$}1
           {'w}{$\omega$}1
           {'state}{$\sigma$}1
           {'w.}{$\forall\omega.\ $}2
           {e1}{e$_1$}1
           {e2}{e$_2$}1
           {...}{\ldots}2
           {\#\#+}{\color{red}}1
           {\#\#-}{\color{black}}1
           {\#\#\#}{{$\leadsto$}}3
}
\lstnewenvironment{code}{\lstset{basicstyle={\linespread{1.02}\sffamily\normalsize}}}{}

\lstMakeShortInline[columns=fullflexible]|

\newcommand{\aside}[1]{\ignorespaces}

\begin{document}

\title{Complete Stream Fusion for Software-Defined Radio}
\subtitle{Extended Abstract}

\author{Tomoaki Kobayashi}
\affiliation{%
  \institution{Tohoku University}
  \country{Japan}}
\email{tomoaki.kobayashi.t3@dc.tohoku.ac.jp}

\author{Oleg Kiselyov}
\orcid{0000-0002-2570-2186}
\affiliation{%
  \institution{Tohoku University}
  \country{Japan}}
\email{oleg@okmij.org}

\begin{abstract}
Software-Defined Radio (SDR) is widely used not only as a practical
application but also as a fitting benchmark of high-performance signal
processing. We report using the SDR benchmark~-- specifically, FM
Radio reception~-- to evaluate the recently developed single-thread stream
processing library strymonas, contrasting it with the synchronous dataflow
system StreamIt. Despite the absence  
of parallel processing or windowing as a core primitive,
strymonas turns out to easily support SDR, offering high
expressiveness and performance, approaching the peak single-core
floating-point performance, sufficient for real-time FM reception.
\end{abstract}
\maketitle

\section{Summary}
\label{s:summary}

\begin{figure}[h]
\includegraphics[width=0.97\columnwidth]{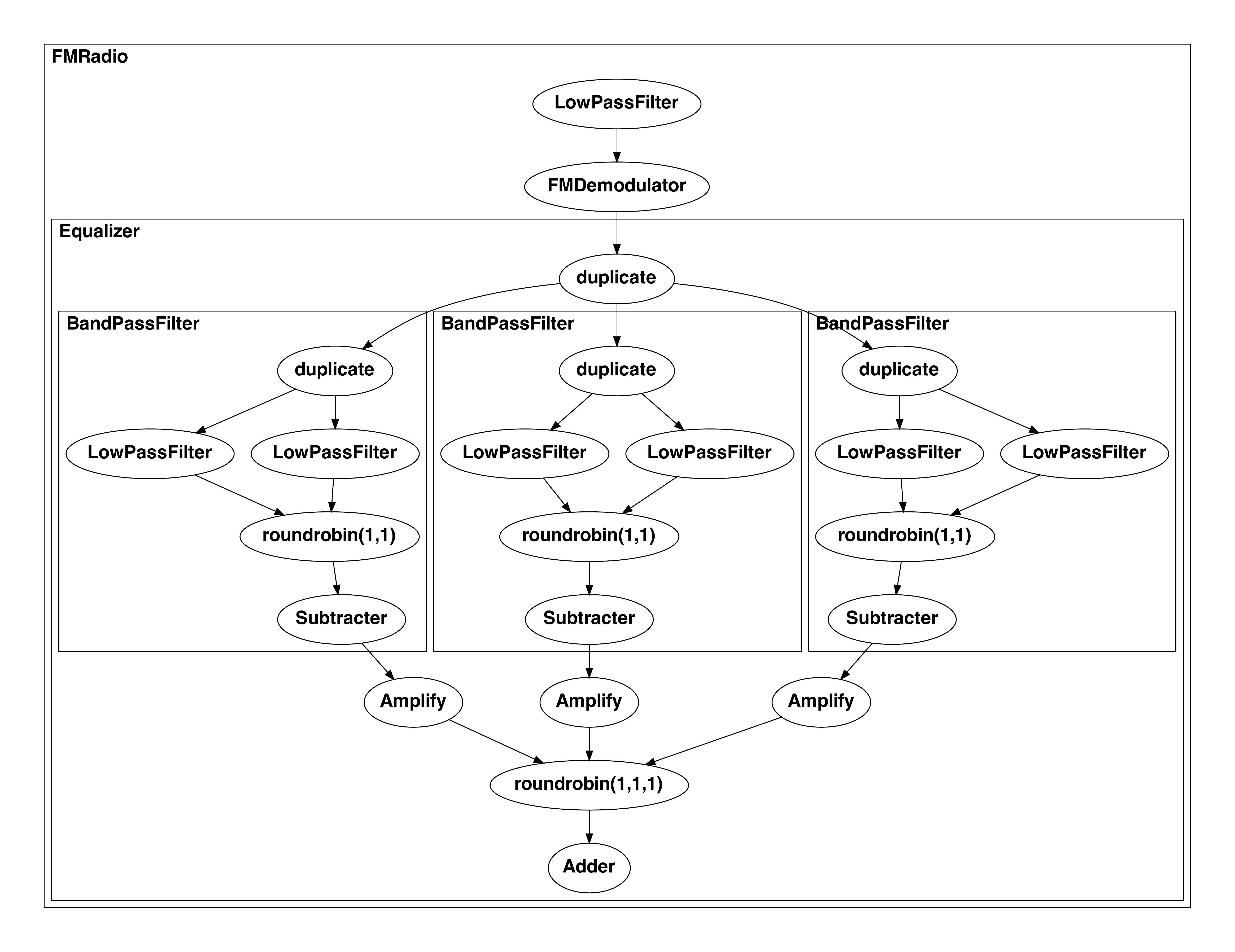}
\caption{FM Radio reception diagram}
\label{f:diagram}
\end{figure}

Software-defined radio (SDR)\footnote{\relax
\url{https://hackrf.readthedocs.io/en/latest/index.html}}
is performing all steps of
radio signal processing (save for the antenna reception or
transmission) not via analog electric circuits but digitally in
software, typically running on an ordinary computer. GNU
Radio\footnote{\relax
\url{https://wiki.gnuradio.org/}} is a
characteristic example. Besides being a widely used application, it
also makes a good benchmark of high-performance signal processing,
used as such in \cite{streamit-thesis,steward_vytiniotis_2015}. An
example is FM Radio reception, diagrammed in Fig.~\ref{f:diagram}
borrowed from \cite{streamit-thesis}.

\subsection{Strymonas}

Strymonas \cite{strymonas-2017} is an embedded DSL for single-thread stream
processing, declaratively assembling stream processing pipelines like
Xmas lights, from operators like |map|, |filter|, |zip|, |flat_map|,
etc.  This work uses the further developed version of strymonas.

The characteristic of strymonas is the static guarantee of
\emph{complete} fusion: if each
operator individually runs without any function
calls and memory allocations however temporary, the entire streaming
pipeline runs without calls and allocations. Thus
strymonas per se introduces not even constant-size intermediary data
structures. An example is |zip_with f|,
defined as |zip >> map (fun (x,y) -> f x y)| where
|>>| is left-to-right function composition. A naive implementation would
construct a tuple in |zip|, to be deconstructed in the subsequent
mapping. In strymonas, no intermediate tuples are constructed.  The
processing loop thus may run without any GC or even stack
allocations~-- of importance to SDR.

Strymonas has several backends to emit code in OCaml, Scala and
C~-- each statically guarantying the generated code compiles
without errors or warnings.  Here we used the C backend, based on
embedding of C in tagless-final style \cite{generating-C}.

In this work we apply strymonas to the FM Radio reception,
Fig.~\ref{f:diagram}. There are immediate problems: Strymonas does not
support split/join (or, duplicate/join) operations apparent in
Fig.~\ref{f:diagram}. Out of the box strymonas also does not support
signal filtering, or any windowing for that matter. These
problems turned out easily surmountable; strymonas, hence, is useful and
performant for SDR.

\section{FM Radio Reception in Strymonas}

Although strymonas does not support windowing out of the box, it
offers enough tools to implement
\begin{code}
make_stream : float cstream -> window stream
\end{code}
that converts a stream of |float| to a window stream, where |window|
is an abstract data type with the operations |reduce| and
|dot| product. Streams in strymonas do not have to be of base
types. As an example, FM demodulation is
\begin{code}
let fmDemodulator (sampRate:float) (max:float) (bndwdth:float)
                    : float cstream -> float cstream =
  let gain = 
      C.float (max *. sampRate /. (bndwdth *. Float.pi)) in
  let (module Win) = Window.make_window C.tfloat 2 0 in
  Win.make_stream
  >> map_raw' (Win.reduce C.( *. ))
  >> map_raw C.(fun e k -> letl (gain *. atan e) k)
\end{code}
The prefix |C| qualifies backend code generation combinators;
|map_raw'| is the mapping operator and |map_raw| is its CPS version, used
for generating a let-binding. We took advantage of OCaml's first-class
modules to provide several implementations of the |window| abstract
type, optimized for a particular window size, sliding/tumbling, etc.
For example, for short windows we use ordinary variables to store past
elements.

To convolve a stream with a given array |arr| we likewise wake a
window stream and reduce the window to a |float| by 
the dot-product with |arr|:
\begin{code}
let apply_filter ?(decimation=0) (arr: float array) 
    : float cstream -> float cstream =
  let coeff = Array.map C.float arr in 
  let taps = Array.length coeff in
  let (module Win) = 
     Window.make_window C.tfloat taps decimation in
  Win.make_stream
  >> map_raw C.(letl (Win.dot tfloat coeff ( +. ) ( *. )))
\end{code}
Then low-pass filtering is the straightforward
\begin{code}
let lowPassFilter (rate:float) (cutoff:float) (taps:int) 
         (decimation:int) : float cstream -> float cstream =
    apply_filter (lpf_coeff rate cutoff taps) ~decimation
\end{code}
where |lpf_coeff rate cutoff taps : float array| computes the coefficients
of the low-pass filter with the given parameters. 

Equalization is splitting the signal into several bands, amplifying by
a band-specific gain and re-combining. This looks, see
Fig.~\ref{f:diagram}, far more complex than |LowPassFilter|. However,
convolutional filters are linear and so we can perform the
equalization on filter coefficients instead. Our equalizer hence is,
like |lowPassFilter|, a mere |apply_filter| operation.

Overall, the FM Radio reception is literally
\begin{code}
lowPassFilter samplingRate cutoffFrequency nTaps 4
>> fmDemodulator samplingRate maxAmplitude bandwidth
>> equalizer samplingRate bands eqCutoff eqGain nTaps
\end{code}
where |samplingRate| (set to 250MHz), |cutoffFrequency| (108MHz),
|bandwidth| (10KHz),
|nTaps| (64), etc. are the standard FM Radio parameters.

\section{Evaluation}
\label{s:eval}

We have verified the correctness of our implementation by writing a naive
(and hence obviously correct) StreamIt interpreter and checking that
the interpretation of all steps of the diagram Fig.~\ref{f:diagram}
gives the same output as the strymonas implementation.  We also
generated a sample FM Radio signal by modulating a sine wave, fed into
our implementation and checked the result by ear.

The C code generated by strymonas for the FM radio reception spans 216
sparsely-filled lines, half of which are filter coefficients. It is
completely vectorized when compiled by GCC.

\begin{figure}[h]
\includegraphics[width=0.97\columnwidth]{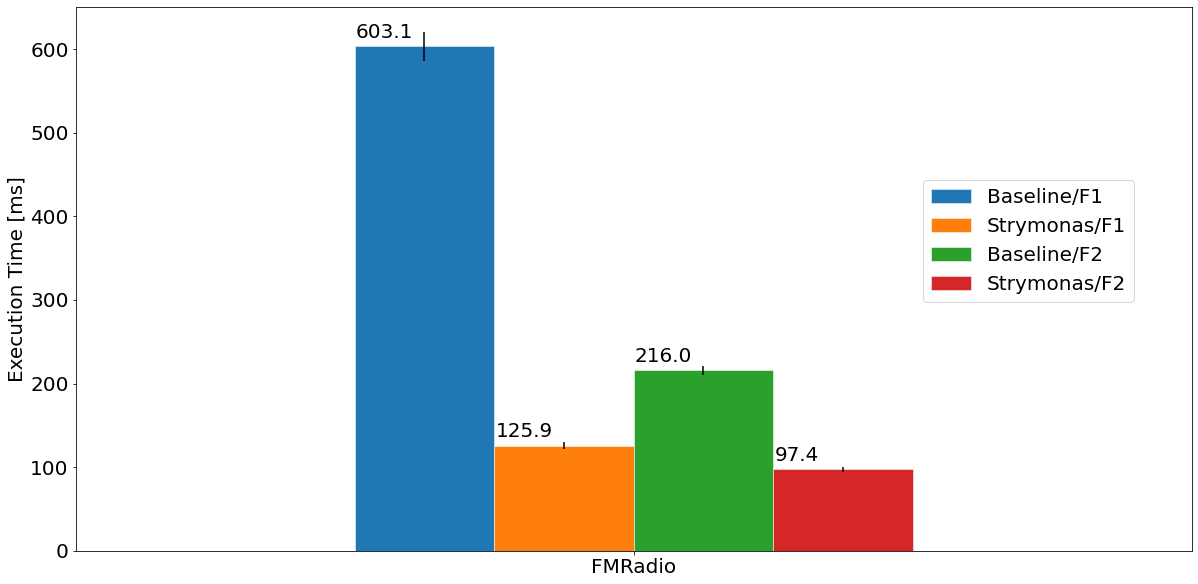}
\caption{Benchmarking against the baseline}
\label{f:bench}
\end{figure}

For performance, we compare with the C reference implementation of FM
radio from the StreamIt
project\footnote{\url{http://groups.csail.mit.edu/cag/streamit/apps/benchmarks/fm/c/fmref.c}}
as the baseline,
with the same benchmark set-up: 1 million synthetic samples, measuring
total processing time.  The evaluation platform is 1.8GHz dualcore
Intel Core i5, 8 GB DDR3 main memory, macOS Big Sur 11.6. The C code
was compiled with GCC 11.2 given the flags "-O3 -march=native
-mfpmath=both -fno-math-errno" (to be called F1) or with
"-ffast-math" added (to be called F2). Fig.~\ref{f:bench} shows
the processing time of 1 million samples, in ms, measured as an
average of 20 runs after 5 warmup runs. Compiled with F2 flags,
strymonas code hence processes in excess of 10M samples/sec, which 
is the average sample rate of the HackRF SDR board.

We also checked memory profile with valgrind. Both the baseline
and our code showed constant memory use throughout the entire
processing, with no heap allocation. 
However, whereas the baseline used 590 Kib of stack, 
our code needed 2KiB (for flags F1) or close to zero (for 
flags F2).

\section{Related work}
\label{s:related}

The FM Radio benchmark was borrowed from StreamIt
\cite{streamit-thesis}: a synchronous dataflow programming language
with static scheduling. In contrast, strymonas is designed for
processing on a single core in a single thread, with all operators
fused rather than scheduled. Although strymonas does not support
split/join, they are not actually necessary in SDR. For example, the
complex equalization in Fig.~\ref{f:diagram} (implemented in StreamIt
as drawn) is reducible to the single ordinary FIR filtering.

GNU Radio includes FM Radio as a standard application.
Unlike GNU Radio, strymonas is typed. Therefore, ill-formed pipelines are
rejected before any code is generated, with error messages describing
ill-fitting operators. The generated code is statically
guaranteed to be well-formed and well-typed, and compiled even without
warnings. Stream elements in strymonas are not limited to base
types: they may be tuples and arbitrary records and objects (such as
|window| seen earlier in the demodulator). The
complete fusion is ensured regardless.

We are currently benchmarking against GNU Radio.

\bibliographystyle{plainnat}
\bibliography{../streams.bib}
\end{document}